\documentclass[prl,nofootinbib,twocolumn,superscriptaddress]{revtex4}
\pdfoutput=1
\usepackage{graphicx,rotating}
\usepackage{hyperref}\usepackage{slashed}
\usepackage{ifpdf}
\usepackage[english]{babel}
\usepackage{amsmath}

\newcommand{\mrm}[1]{\mbox{\rm #1}}
\newcommand{\be}{\begin{equation}}
\newcommand{\ee}{\end{equation}}
\newcommand{\br}{\begin{eqnarray}}
\newcommand{\bea}{\begin{eqnarray}}
\newcommand{\eea}{\end{eqnarray}}
\newcommand{\er}{\end{eqnarray}}
\newcommand{\ba}{\begin{array}}
\newcommand{\ea}{\end{array}}
\newcommand{\bi}{\begin{itemize}}
\newcommand{\ei}{\end{itemize}}
\newcommand{\bn}{\begin{enumerate}}
\newcommand{\en}{\end{enumerate}}
\newcommand{\bc}{\begin{center}}
\newcommand{\ec}{\end{center}}

\newcommand{\no}{\nonumber}
\newcommand{\nn}{\nonumber\\}

\newcommand{\Eq}[1]{Eq.~(\ref{#1})}
\newcommand{\rfn}[1]{(\ref{#1})}

\newcommand{\gsim}{\lower.7ex\hbox{$\;\stackrel{\textstyle>}{\sim}\;$}}
\newcommand{\lsim}{\lower.7ex\hbox{$\;\stackrel{\textstyle<}{\sim}\;$}}

\newcommand{\hc}[1]{#1^{\dagger}} % Hermitian conjugate

\def\mysection#1{{\bf #1.} }

\begin{document}
%\draft

%\preprint

\title{\bf 
Matter parity as the origin of scalar Dark Matter
}

\author{ Mario Kadastik}
\author{ Kristjan Kannike}
\author{ Martti Raidal}

\affiliation{National Institute of Chemical Physics and Biophysics, 
Ravala 10, Tallinn 10143, Estonia
}

%\date{\today}
%\pacs{}

\vspace*{1.in}

\begin{abstract}

We extend the concept of matter parity $P_M=(-1)^{3(B-L)}$ to non-supersymmetric theories and argue that 
$P_M$ is the natural explanation  to the existence of  Dark Matter  of the Universe. We show that 
the non-supersymmetric Dark Matter must 
be contained in scalar $\bf 16$ representation(s) of $SO(10),$ thus the unique low energy Dark Matter candidates are
$P_M$-odd complex scalar singlet(s) $S$ and inert  scalar doublet(s) $H_2.$ We have calculated the thermal relic Dark Matter  abundance of the model and shown that its minimal form may be testable at LHC via the SM Higgs boson decays $H_1\to DM\;DM.$ The PAMELA anomaly can be explained with the decays $DM\to \nu l W$ induced via  seesaw-like operator which is additionally 
suppressed by Planck scale.  Because the SM fermions are odd under matter parity too, the DM sector is  just our scalar  relative.

\end{abstract}

\vskip -3.5cm

\maketitle

\mysection{Introduction}
While the existence of Dark Matter (DM) of the Universe is now established without doubt~\cite{Komatsu:2008hk},
its origin, nature and properties remain obscured. Any well motivated theory beyond the standard model (SM)
must explain what constitutes  the DM and why those DM particles are stable. 
In most  popular models beyond the SM, such as the minimal supersymmetric SM, 
additional discrete $Z_2$ symmetry is imposed by hand to ensure the stability of the lightest $Z_2$-odd particle. 
There is no known general physics principle for the origin of DM which could  discriminate between the 
proposed DM models.

In this Letter we propose that  there actually might exist such a common physics 
principle for the theories of DM.  It follows from  the underlying unified symmetry group for
all matter fields in grand unified theories (GUTs) and does not require supersymmetry.
One can classify all matter fields  in Nature under the 
discrete remnant of the GUT symmetry group which is nothing but the matter parity $P_M.$
Thus the existence of DM might be a general property of Nature rather than an accidental outcome
of  some particular model. As a general result, there is no ``dark world" decoupled from us, rather we are 
part of it as the SM fermions are also odd under the matter parity $P_M.$

We argue that, assuming  $SO(10)$~\cite{Fritzsch:1974nn} to be the GUT symmetry group,
the discrete center $Z_n$ of $U(1)_X\in SO(10)$ remains unbroken.  
For the simplest case, $n=2,$ the GUT symmetry breaking chain  $SO(10)\to SU(5)\times P_M$ 
implies that all the fermion and scalar fields of the GUT theory, including the SM particles plus the right-handed neutrinos $N_i,$ 
carry well defined discrete quantum  numbers which are uniquely determined by their original representation 
under $SO(10).$  We show that non-supersymmetric DM candidates can come only
from $\bf 16$ scalar representations of  $SO(10),$  and the unique  low energy DM fields  are new
$SU(2)_L\times U(1)_Y$ 
 $P_M$-odd scalar doublet(s) $H_2$~\cite{id} and singlet(s) $S$~\cite{rs,cs}.

We formulate and study the minimal matter parity induced phenomenological
DM model which contains one inert doublet $H_2$ and one complex singlet $S.$
We show that the observed DM thermal freeze-out abundance can be achieved for wide range of model 
parameters. 
We also show that the  PAMELA~\cite{pam} and ATIC~\cite{atic} anomalies in 
$e^+/(e^-+e^+)$ and $e^-+e^+$ cosmic ray fluxes
can be explained by  DM decays via $d=6$~\cite{Arvanitaki:2008hq} operators. In our case the
 Planck scale suppressed
$P_M$-violating seesaw-like operator is of the form 
$ m/(\Lambda_N M_P) LLH_1H_2,$   where $m/M_P$ is  $P_M$-violating heavy neutrino mixing.
In this model the SM Higgs boson $H_1$ is the portal~\cite{Patt:2006fw} to the DM.
We show that for well motivated model parameter the DM abundance predicts the decay $H_1\to DM\;DM, $
which allows to test the model at LHC~\cite{Ball:2007zza}.

\mysection{Matter parity as the origin of DM} 
The prediction of $SO(10)$ GUT is that the  fermions of every  generation  form one $SO(10)$ multiplet $\bf 16_i,$ $i=1,2,3.$ 
This is in a perfect agreement with experimental data as there exist 15 SM fermions per generation 
plus right-handed $N_i$ for the seesaw mechanism~\cite{seesaw}. 
Assuming  $SO(10)$ GUT,  the first step in the group theoretic branching rule for the GUT symmetry breaking,
\bea
SO(10) \to SU(5) \times U(1)_X \to SU(5)\times Z_2,
\label{break}
\eea
implies that every $SU(5)$   matter multiplet~\cite{Georgi:1974sy}  and $N_i$ carry an additional uniquely  defined
quantum number under the   $U(1)_X$ symmetry.  
The $U(1)_X$ symmetry can be further broken to its discrete subgroup $Z_n$ by  an order
parameter carrying $n$ charges of $X$ \cite{Krauss:1988zc,Martin:1992mq}.  
The simplest case $Z_2,$ which allows for the seesaw mechanism induced by the  heavy neutrinos $N_i$ \cite{seesaw}, 
 yields the new parity $P_X$ with the   field transformation $\Phi\to \pm\Phi .$ 
Therefore, at the electroweak scale after $SU(5)$ symmetry breaking, 
the actual SM symmetry group becomes $SU(2)_L\times U(1)_Y \times P_X.$
The discrete remnant of the GUT symmetry group, $P_X,$ implies the existence of stable  DM.

Under  Pati-Salam charges  $B-L$ and $T_{3R}$ the
$X$-charge is decomposed as 
\be
X=3 (B-L)+4 T_{3R},
\ee
while the orthogonal combination, the SM hypercharge $Y,$
is gauged in  $SU(5).$ 
Because  $X$ depends on $4 T_{3R}$ which is always an even integer for  $T_{3R}=1/2,\;1, ...,$ 
the $Z_2$ $X$-parity of a multiplet 
is determined by $3 (B-L)\; \mrm{mod 2}.$ 
Therefore one can write
\bea
P_{X}= P_M=(-1)^{3(B-L)},
\label{P}
\eea
and identify $P_X$ with the well known matter parity~\cite{Farrar:1978xj}, 
 which is equivalent to $R$-parity  in supersymmetry. 
While $U(1)_X,$ $X=5(B-L)-2Y,$ has been used to discuss and to forbid proton decay  
operators \cite{Wilczek:1979et}, so far
the parity \rfn{P} has been associated only with SUSY phenomenology.

Due to \Eq{break} a definite matter parity $P_M$ is the general intrinsic property of
every matter multiplet. 
The decomposition of $\bf 16$ of
$SO(10)$ under  \rfn{break}  is ${\bf 16} ={\bf 1}^{16}(5)+{\bf \bar 5}^{16} (-3) + {\bf 10}^{16} (1),$
where the $U(1)_X$ quantum numbers of the $SU(5)$ fields are given in brackets. 
This implies that under the matter parity all the fields ${\bf 10}^{16} ,$ ${\bf \bar 5}^{16} ,$ ${\bf 1}^{16} $ 
are odd. At the same time, all other fields coming from small $SO(10)$ representations, 
${\bf 10} ,$ ${\bf 45} ,$ ${\bf 54} ,$ ${\bf 120} $  and ${\bf 126} ,$ are predicted to be even under $P_M.$
Thus the SM fermions belonging to ${\bf 16} _i$ are all $P_M$-odd while the SM Higgs boson doublet is $P_M$-even
because it is embedded into  ${\bf 5}^{10}$ and/or ${\bf \bar 5}^{10},$ and  ${\bf 10} ={\bf 5}^{10}(-2)+{\bf \bar 5}^{10} (2). $
Although $B-L$ is broken in nature by heavy neutrino Majorana masses, $(-1)^{3(B-L)}$ is respected by interactions of 
{\it all} matter fields.

As there is no DM candidate in the SM, we have to extend the particle content of the model 
by adding new $SO(10)$ multiplets. The choice is {\it unique} as only ${\bf 16}$ contains $P_M$-odd particles.  
Adding a new fermion  $\bf 16$ is equivalent to adding a new generation, and this does not give DM.
Thus  we have only one possibility, the scalar(s)  $\bf 16$ of  $SO(10).$
Because DM must be electrically neutral,  
$\bf 16$ contains only two DM candidates.  Under  $SU(2)_L\times U(1)_Y$ those are 
 the complex singlet $S={\bf 1}^{16}$ and the inert doublet 
$H_2\in {\bf \bar 5}^{16}.$

\mysection{DM predictions of the minimal model}
GUT symmetry groups are known to be very useful for classification of particle quantum numbers,
and this is sufficient for  predicting the DM candidates. Unfortunately 
GUTs fail, at least in their minimal form,  to predict correctly coupling constants between matter fields.
Therefore we cannot trust GUT model building for predicting details of DM phenomenology.
Instead we study {\it phenomenological  low-energy} Lagrangian for the SM Higgs $H_1$ and the
$P_M$-odd scalars $S$ and $H_2,$
\bea
{V} &=&
 -\mu_1^{2} \hc{H_1} H_1 + \lambda_{1} (\hc{H_1} H_1)^{2} + \mu_{S}^{2} \hc{S} S + \lambda_{S} (\hc{S} S)^{2} 
 \nn
& +& \lambda_{S H_{1}}( \hc{S} S) (\hc{H_1} H_1) + \mu_{2}^{2} \hc{H_2} H_2 + \lambda_{2} (\hc{H_2} H_2)^{2} \nn
&+ & \lambda_{3} (\hc{H_1} H_1) (\hc{H_2} H_2) + \lambda_{4} (\hc{H_1} H_2) (\hc{H_2} H_1) \nn
&+& \frac{\lambda_{5}}{2} \left[ (\hc{H_1} H_2)^{2} + (\hc{H_2} H_1)^{2} \right]  
+   \frac{b_{S}^2}{2} \left[ S^{2} + (S^\dagger)^{2} \right]   \label{V} \\
&+& \lambda_{S H_2} (\hc{S} S) (\hc{H_2} H_2) + \frac{\mu_{S H}}{2}  \left[S^\dagger \hc{H_1} H_2 + S \hc{H_2} H_1 \right],\no
\eea
which respects $H_1 \to H_1$ and $S \to -S, H_2 \to -H_2.$ 
The doublet terms alone form the inert doublet 
model~\cite{id}.   Following  Ref.~\cite{cs}, to ensure $\langle S \rangle=0,$ 
we allow only the soft mass terms $b_S,$ $\mu_{SH}$ and the $\lambda_5$ term to break the internal $U(1)$ of the odd scalars. 
Thus the singlet terms in \rfn{V} alone form the model A2 of~\cite{cs}. 
The two models mix via $\lambda_{SH},$ $\mu_{SH}$ terms.
Notice that mass-degenerate scalars are strongly constrained as DM candidates by direct searches for DM.
The $\lambda_5,$ $b^2_S$ and $\mu_{SH}$ terms in \Eq{V} are crucial for lifting the mass degeneracies.

\begin{figure}[t]
\includegraphics[width=0.38\textwidth]{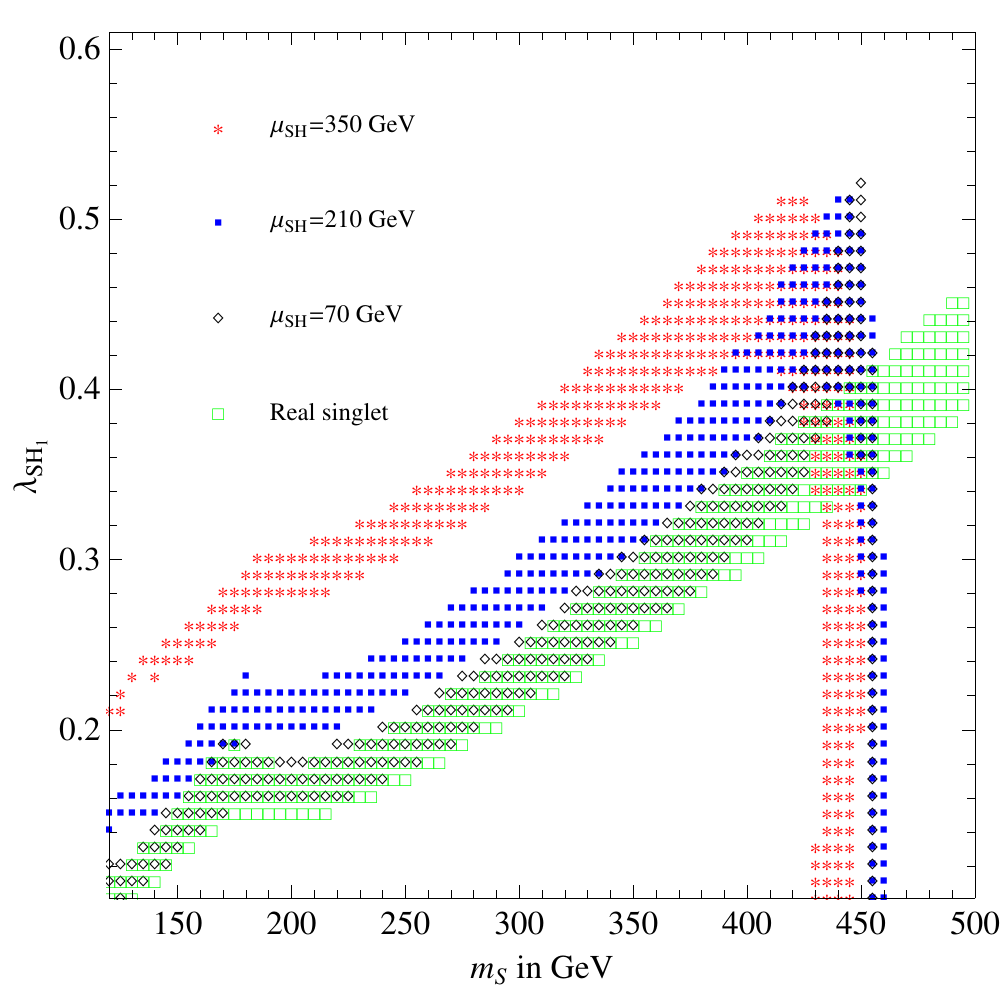}
\caption{ Allowed $3\sigma$ regions 
for predominantly singlet DM in  $(m_S, \;\lambda_{SH_1})$ plane for 
$b_S=5$GeV, $m_{H_0}=450$  GeV. 
}
\label{fig1}
\end{figure}

We stress that our model of DM is based on the particle quantum numbers
and does not rely on numerology. However, the phenomenological studies of the model 
necessarily rise questions such as the gauge coupling unification.  
The one-loop $\beta$-functions for gauge couplings  $g$, $g'$ and $g_{3}$ are given by
$  \beta_{g'} = 7 g^{\prime 3} $, $  \beta_{g} = -3 g^{3}$ and $\beta_{g_{3}} = -7 g^{3}.$
Based solely on the running due to those beta functions, 
we identify the unification scale $2\times 10^{16}$~GeV by thesolution for $g_2=g_3.$
The exact values of gauge couplings at $M_G$ are given by $g_1= \sqrt{5/3} g' = 0.58,$ $g_2=g_3=0.53.$
The precision of unification of all three gauge couplings in our model is better than in the SM because of the existence of an
extra scalar doublet. 
 We assume that an exact unification can be achieved due to the GUT thresholds corrections in full 
$SO(10)$ theory which we cannot estimate because the details of GUT symmetry breaking are not known~\cite{Hall:1980kf}.
In the minimal model with one extra doublet the required change of $g_1$ at the GUT
 scale due to the threshold corrections is 10\%.  If, for example, there is one DM scalar multiplet for each 
 generation of fermions, the required threshold corrections are smaller, at the level of 4\%.

In the following we assume that DM is a thermal relic and calculate its abundance using 
 MicrOMEGAs package~\cite{micromegas}. The DM  interactions \rfn{V} were calculated using  FeynRules 
package~\cite{Christensen:2008py}. 
To present numerical examples
we fix  the  doublet parameters following Ref.~\cite{LopezHonorez:2006gr} as  $m_{A_0} - m_{H_0} =  
10$~GeV, $m_{H^\pm} - m_{H_0} = 50$~GeV and treat $m_{H_0}$ and $\mu_2$ as  free parameters.
For predominantly singlet DM we present in Fig.~\ref{fig1} the allowed $3\sigma$  regions 
in the  $m_S^2= \mu_{S}^{2} + \lambda_{SH_1} v^2/2-b_S^2$ and 
 $\lambda_{SH_1}$  plane  for  $b_S=5$~GeV,  $m_{H_0}=450$~GeV and the values of  $\mu_{SH}$ as indicated in the figure. 
For comparison we also plot the corresponding
prediction of the real scalar model (light green band). 
For those parameters the observed DM abundance can be obtained for $m_S<m_{H_0}.$ 
Due to the mixing parameter $\mu_{SH},$ 
 a large region in the  $(m_S, \;\lambda_{SH_1})$  plane becomes   viable.

\begin{figure}[t]
\includegraphics[width=0.5\textwidth]{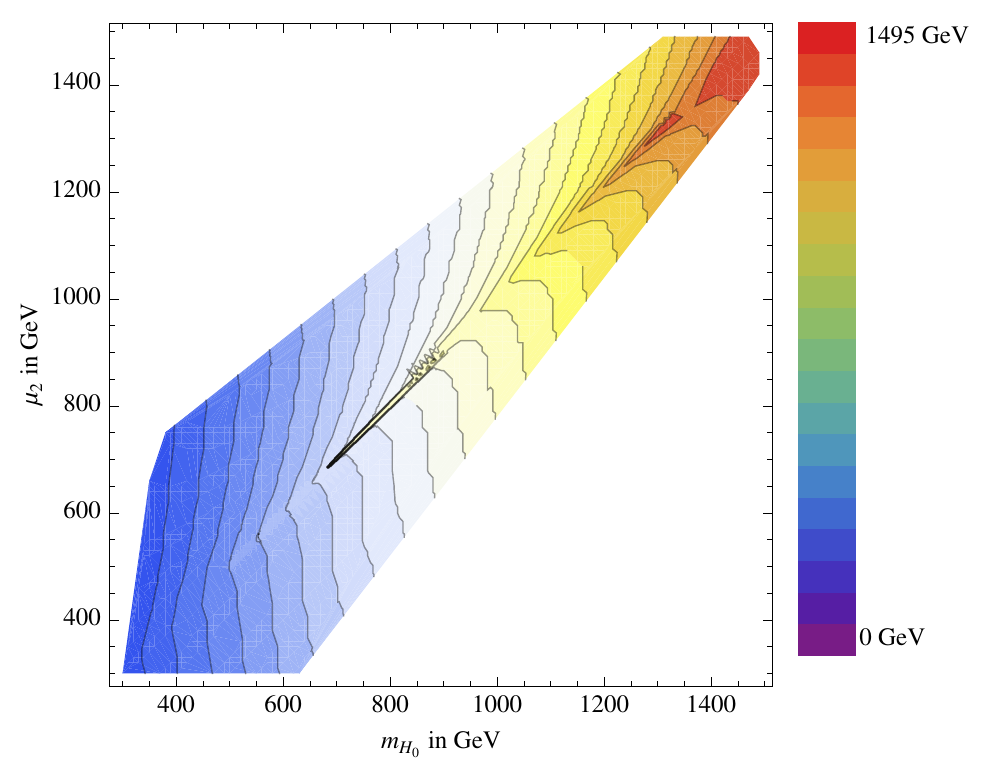}
\caption{Allowed $(m_{H_0},\;\mu_2)$ parameter space for $\mu_{SH}=0$ and different
values of  $m_S$ represented by the color code. }
\label{fig2}
\end{figure}
To study DM dependence on  doublet  parameters we present in Fig.~\ref{fig2} the   $(m_{H_0},\;\mu_2)$ parameter space for which
the observed  DM abundance can be obtained.  Values of the singlet mass are presented by the colour code and we take 
 $\mu_{SH}=0,$ $b_S=5$ GeV.
Without singlet $S,$ in the inert doublet model~\cite{LopezHonorez:2006gr}, 
the allowed parameter space is the narrow region
on the diagonal of Fig.~\ref{fig2} starting at $m_{H_0}\approx 670$~GeV. 
In our model much larger parameter space becomes available.

\mysection{PAMELA, ATIC and FERMI data}
PAMELA satellite  has observed steep rise of $e^+/(e^-+e^+)$ cosmic ray flux with energy and no excess in $\bar p/p$ 
ratio~\cite{pam}.
ATIC experiment  claims a peak in $e^-+e^+$ cosmic ray flux around 700~GeV~ \cite{atic}, 
a claim that will be checked by FERMI satellite soon.
To explain  the cosmic $e^+$ excess with annihilating DM requires enhancement of the annihilation cross section by a
factor $10^{3-4}$ compared to what is predicted for a thermal relic. 
Non-observation of photons associated with annihilation~\cite{Bertone:2008xr}
and the absence of hadronic annihilation modes~\cite{Cirelli:2008pk} 
constrains this scenario very strongly. 
However, the PAMELA anomaly can also be explained with decaying 
thermal relic DM  with lifetime $10^{26}~\mathrm{s}$~\cite{dmdecay}, 3-body decays in our case.

In our scenario the global $Z_2$ matter parity can be broken by Planck scale effects~\cite{Krauss:1988zc}.
If there exists, at Planck scale, 
a $SO(10)$ fermion singlet $N',$  its mixing with the $SU(5)$ $P_M$-odd singlet neutrinos $N$ via
a mass term $mNN'$ breaks  $P_M$ explicitly but softly. 
The exchange of $N$ now induces also a seesaw-like~\cite{seesaw}  operator 
\bea
\frac{\lambda_N}{M_N}\frac{m}{M_{P}} LL H_1 H_2 \to 10^{-30} ~\mrm{GeV}^{-1}~\nu l^- W^+ H_2^0,
\eea
where we have taken $\lambda_N\sim 1,$ $M_N\sim 10^{14}$~GeV
and $m\sim v\sim 100$~GeV. Such a small effective Yukawa coupling explains the long DM lifetime $10^{26}~\mathrm{s}$.

\mysection{LHC phenomenology}
In our scenario the DM  couples to the SM only via the Higgs
boson couplings \Eq{V}. Therefore, discovering $\sim 1$~TeV DM particles at LHC is very challenging. 
However,  if DM is relatively light  the SM Higgs decays $H_1\to DM\, DM$ become kinematically allowed and
the SM Higgs branching ratios are strongly affected. Such a scenario has been studied by LHC 
experiments~\cite{Ball:2007zza} and can be used to discover light scalars.

\begin{figure}[t]
\includegraphics[width=0.3\textwidth]{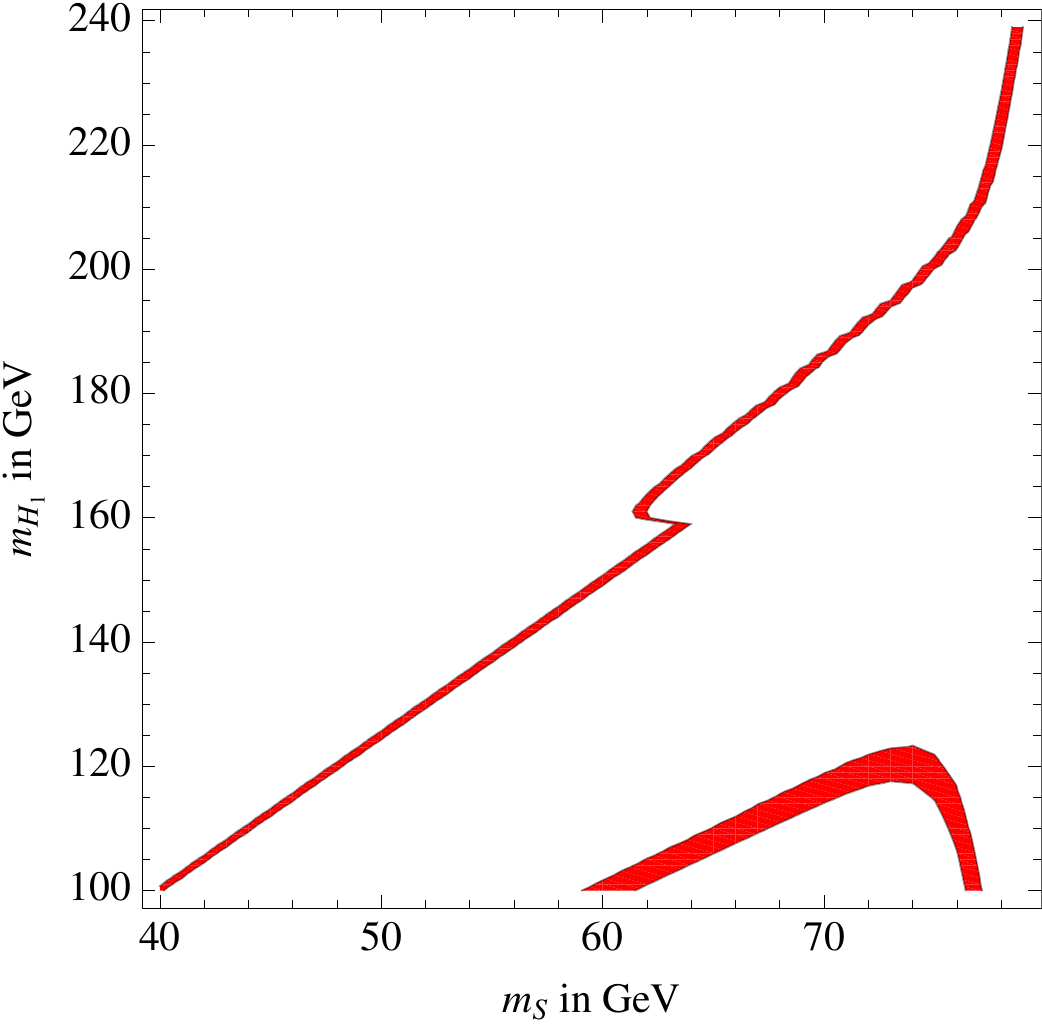}
\caption{ Allowed $3\sigma$ regions in the singlet DM and SM Higgs boson mass  plane for $\mu_S=0$
and $b_S=5$~GeV.}
\label{fig3}
\end{figure}
In our model such a scenario is realized for  $\mu_S=0,$ small $b_S\ll v$  and heavy doublet. 
In this case the DM is predominantly split singlet and, in addition, the DM abundance relates the DM mass 
$m^2_S\approx \lambda_{SH_1}v^2/2-b_S^2$ to the SM
Higgs boson mass $m_{H_1},$ as seen in Fig.~\ref{fig3}. 
For $m_{H_1}=120$~GeV, $b_S=5$~GeV
we predict $m_{S}=48$~GeV with the Higgs branching ratios
$BR(H_1\to b\bar b + c\bar c +\tau\bar \tau)=14.2\%, $
$BR(H_1\to DM\,DM)=42.4\%$ and $BR(H_1\to S_2\, S_2)=42.4\%.$ 
The second heaviest singlet $S_2$ with the mass $m^2_{S_2}\approx \lambda_{SH_1}v^2/2+b_S^2$
decays via the SM Higgs exchange to $S_2\to DM\,\mu\,\bar\mu$ or $S_2\to DM\,c\,\bar c$
with almost equal branching ratios. 
Thus the SM Higgs boson decay modes are very strongly modified. This makes the $H_1$ discovery 
more difficult at LHC but, on the other hand,  allows the scenario to be   tested 
via the Higgs portal~\cite{Patt:2006fw}.

\mysection{Conclusions}
We have extended the concept of $Z_2$ matter parity, $P_{M}=(-1)^{3(B-L)},$  to 
non-supersymmetric GUTs and argued that $P_M$  gives the natural origin of  DM of the Universe.
Assuming that $SO(10)$ is the GUT symmetry group,
the matter parity of all  matter multiplets is determined by their $U(1)_X$ charge under  \Eq{break}.  
Consequently, the non-supersymmetric DM must be contained in the scalar representation $\bf 16$  of $SO(10).$   
 This implies that the theory of DM becomes completely predictive and the only possible low energy DM candidates are
 the $P_M$-odd   scalar singlet(s)  $S$ and doublet(s) $H_2.$ 
 We have calculated the DM abundances
 in the minimal DM  model and shown that  it has a chance to be tested at LHC via Higgs portal.
 Planck-suppressed  $P_M$ breaking effects may occur in the heavy neutrino sector
 leading to decays $DM\to \nu l W$ which can explain the PAMELA  and FERMI anomalies.

Our main conclusion is that there is nothing unusual   in the DM which
 is just  scalar relative of the SM fermionic matter. Although $B-L$ is broken
 in Nature by heavy neutrino Majorana masses, $(-1)^{3(B-L)}$ is respected by interactions of all matter fields
 implying stable scalar DM.

\vskip 0.3 cm

\noindent {\bf Acknowledgment.}
We thank S. Andreas, J. van der Bij, M. Cirelli, Y. Kajiyama, M. Kachelriess, M. Picariello,  A. Romanino, and A. Strumia for discussions.
This work was supported by the ESF Grant 8090 and
by EU  FP7-INFRA-2007-1.2.3 contract No 223807.


\begin{thebibliography}{99}

%\cite{Komatsu:2008hk}
\bibitem{Komatsu:2008hk}
  E.~Komatsu {\it et al.}  [WMAP Collaboration],
  %``Five-Year Wilkinson Microwave Anisotropy Probe (WMAP\altaffilmark 1 )
  %Observations:Cosmological Interpretation,''
  Astrophys.\ J.\ Suppl.\  {\bf 180}, 330 (2009)
  [arXiv:0803.0547].
  %%CITATION = APJSA,180,330;%%
  
    %\cite{Fritzsch:1974nn}
\bibitem{Fritzsch:1974nn}
  H.~Fritzsch and P.~Minkowski,
  %``Unified Interactions Of Leptons And Hadrons,''
  Annals Phys.\  {\bf 93}, 193 (1975).
  %%CITATION = APNYA,93,193;%%

     
    \bibitem{id}
%\cite{Deshpande:1977rw}
%\bibitem{Deshpande:1977rw}
  N.~G.~Deshpande and E.~Ma,
  %``Pattern Of Symmetry Breaking With Two Higgs Doublets,''
  Phys.\ Rev.\  D {\bf 18}, 2574 (1978);
  %%CITATION = PHRVA,D18,2574;%%
%\cite{Barbieri:2006dq}
%\cite{Ma:2006km}
%\bibitem{Ma:2006km}
  E.~Ma,
  %``Verifiable radiative seesaw mechanism of neutrino mass and dark matter,''
  Phys.\ Rev.\  D {\bf 73}, 077301 (2006);
  %%CITATION = PHRVA,D73,077301;%%
%\bibitem{Barbieri:2006dq}
  R.~Barbieri, L.~J.~Hall and V.~S.~Rychkov,
  %``Improved naturalness with a heavy Higgs: An alternative road to LHC
  %physics,''
  Phys.\ Rev.\  D {\bf 74}, 015007 (2006)
  [arXiv:hep-ph/0603188].
  %%CITATION = PHRVA,D74,015007;%%



\bibitem{rs}
%\cite{McDonald:1993ex}
%\bibitem{McDonald:1993ex}
  J.~McDonald,
  %``Gauge Singlet Scalars as Cold Dark Matter,''
  Phys.\ Rev.\  D {\bf 50}, 3637 (1994);
  %%CITATION = PHRVA,D50,3637;%%
%\cite{Burgess:2000yq}
%\bibitem{Burgess:2000yq}
  C.~P.~Burgess, M.~Pospelov and T.~ter Veldhuis,
  %``The minimal model of nonbaryonic dark matter: A singlet scalar,''
  Nucl.\ Phys.\  B {\bf 619}, 709 (2001);
  %%CITATION = NUPHA,B619,709;%%
  V.~Barger  {\it et al.,}
    %``LHC Phenomenology of an Extended Standard Model with a Real Scalar
  %Singlet,''
  Phys.\ Rev.\  D {\bf 77}, 035005 (2008).
   
  \bibitem{cs}
%\cite{Barger:2008jx}
%\bibitem{Barger:2008jx}
  V.~Barger  {\it et al.,}
  %``Complex Singlet Extension of the Standard Model,''
  Phys.\ Rev.\  D {\bf 79}, 015018 (2009).
  %%CITATION = PHRVA,D79,015018;%%

    

\bibitem{pam}
%\cite{:2008zzr}
%\cite{Adriani:2008zr}
%\bibitem{Adriani:2008zr}
  O.~Adriani {\it et al.}  [PAMELA Collaboration],
  %``Observation of an anomalous positron abundance in the cosmic radiation,''
  arXiv:0810.4995 [astro-ph];
  %%CITATION = ARXIV:0810.4995;%%
%\cite{Adriani:2008zq}
%\bibitem{Adriani:2008zq}
  O.~Adriani {\it et al.},
 %  ``A new measurement of the antiproton-to-proton flux ratio up to 100 GeV in
  %the cosmic radiation,''
  Phys.\ Rev.\ Lett.\  {\bf 102}, 051101 (2009).
    %%CITATION = PRLTA,102,051101;%%

\bibitem{atic}
  J.~Chang {\it et al.},
  %``An Excess Of Cosmic Ray Electrons At Energies Of 300.800 Gev,''
  Nature {\bf 456}, 362 (2008).
  %%CITATION = NATUA,456,362;%%

%\cite{Arvanitaki:2008hq}
\bibitem{Arvanitaki:2008hq}
  A.~Arvanitaki, S.~Dimopoulos, S.~Dubovsky, P.~W.~Graham, R.~Harnik and S.~Rajendran,
  %``Astrophysical Probes of Unification,''
  arXiv:0812.2075 [hep-ph].
  %%CITATION = ARXIV:0812.2075;%%


%\cite{Patt:2006fw}
\bibitem{Patt:2006fw}
  B.~Patt and F.~Wilczek,
  %``Higgs-field portal into hidden sectors,''
  arXiv:hep-ph/0605188.
  %%CITATION = HEP-PH/0605188;%%

%\cite{Ball:2007zza}
\bibitem{Ball:2007zza}
  G.~L.~Bayatian {\it et al.}  [CMS Collaboration],
  %``CMS technical design report, volume II: Physics performance,''
  J.\ Phys.\ G {\bf 34}, 995 (2007).
  %%CITATION = JPHGB,G34,995;%%

 
  
\bibitem{seesaw} 
 P.~Minkowski, %``Mu $\to$ E Gamma At A Rate Of One Out Of 1-Billion Muon Decays?,''
Phys.\ Lett.\ B {\bf 67}, 421 (1977);
T.~Yanagida, 
%``Horizontal symmetry and masses of neutrinos,''
in {\it  Baryon Number of the Universe and Unified Theories, } 
Tsukuba, Japan, 13-14 Feb 1979;
M.~Gell-Mann, P.~Ramond and R.~Slansky,
% ``Complex Spinors And Unified Theories,''
in {\it Supergravity,} P. van Nieuwenhuizen and D.Z. Freedman (eds.), North Holland Publ. Co., 1979;
 S.~L.~Glashow, %``The Future Of Elementary Particle Physics,''
NATO Adv.\ Study Inst.\ Ser.\ B Phys.\  {\bf 59} (1979) 687;
R.~N.~Mohapatra and G.~Senjanovic, %``Neutrino mass and spontaneous parity nonconservation,''
Phys.\ Rev.\ Lett.\  {\bf 44} (1980) 912.

  
   %\cite{Georgi:1974sy}
\bibitem{Georgi:1974sy}
  H.~Georgi and S.~L.~Glashow,
  %``Unity Of All Elementary Particle Forces,''
  Phys.\ Rev.\ Lett.\  {\bf 32}, 438 (1974).
  %%CITATION = PRLTA,32,438;%%

  
  
  
  %\cite{Krauss:1988zc}\cite{Martin:1992mq}\cite{Farrar:1978xj}
\bibitem{Krauss:1988zc}
  L.~M.~Krauss and F.~Wilczek,
  %``Discrete Gauge Symmetry in Continuum Theories,''
  Phys.\ Rev.\ Lett.\  {\bf 62}, 1221 (1989).
  %%CITATION = PRLTA,62,1221;%%
  
  %\cite{Martin:1992mq}
\bibitem{Martin:1992mq}
  S.~P.~Martin,
  %``Some simple criteria for gauged R-parity,''
  Phys.\ Rev.\  D {\bf 46}, 2769 (1992).
  %%CITATION = PHRVA,D46,2769;%%
  
  %\cite{Farrar:1978xj}
\bibitem{Farrar:1978xj}
  G.~R.~Farrar and P.~Fayet,
  %``Phenomenology Of The Production, Decay, And Detection Of New Hadronic
  %States Associated With Supersymmetry,''
  Phys.\ Lett.\  B {\bf 76}, 575 (1978);
  %%CITATION = PHLTA,B76,575;%%
  %\cite{Dimopoulos:1981zb}
%\bibitem{Dimopoulos:1981zb}
  S.~Dimopoulos and H.~Georgi,
  %``Softly Broken Supersymmetry And SU(5),''
  Nucl.\ Phys.\ B {\bf 193}, 150 (1981);
  %%CITATION = NUPHA,B193,150;%%
    L.~Ibanez and G.~Ross,
  %``Discrete Gauge Symmetries And The Origin Of Baryon And Lepton Number
  %Conservation In Supersymmetric Versions Of The Standard Model,''
  Nucl.\ Phys.\  B {\bf 368}, 3 (1992).
  %%CITATION = NUPHA,B368,3;%%
  
  %\cite{Wilczek:1979et}
\bibitem{Wilczek:1979et}
  F.~Wilczek and A.~Zee,
  %``Conservation Or Violation Of B-L In Proton Decay,''
  Phys.\ Lett.\  B {\bf 88}, 311 (1979);
  %%CITATION = PHLTA,B88,311;%%
 N.~Sakai and T.~Yanagida,
  %``Proton Decay In A Class Of Supersymmetric Grand Unified Models,''
  Nucl.\ Phys.\  B {\bf 197}, 533 (1982).
  %%CITATION = NUPHA,B197,533;%%
  
  
%\bibitem{footnote1}
%The $\lambda_5$ term is generated by SO(10) invariant operator
%${\bf 16}\,{\bf 10}\,{\bf 16}\,{\bf 10}\,{\bf {\overline{126}}}/M_P$ 
%after the breaking of left-right symmetry by $\langle{\bf {\overline{126}}}\rangle.$
%We predict $\lambda_5\sim M_{GUT}/M_P\sim 10^{-2}.$



\bibitem{Hall:1980kf}
  L.~J.~Hall,
  %``Grand Unification Of Effective Gauge Theories,''
  Nucl.\ Phys.\  B {\bf 178}, 75 (1981);
  %%CITATION = NUPHA,B178,75;%%
  V.~V.~Dixit and M.~Sher,
  %``THE FUTILITY OF HIGH PRECISION SO(10) CALCULATIONS,''
  Phys.\ Rev.\  D {\bf 40}, 3765 (1989).
  %%CITATION = PHRVA,D40,3765;%%




\bibitem{micromegas}
 %\bibitem{Belanger:2006is}
   G.~Belanger, F.~Boudjema, A.~Pukhov and A.~Semenov,
   %`micrOMEGAs2.0: A program to calculate the relic density of dark  
%matter  in
   %a generic model,''
   Comput.\ Phys.\ Commun.\  {\bf 176} (2007) 367.
%\cite{Christensen:2008py}

\bibitem{Christensen:2008py}
  N.~D.~Christensen and C.~Duhr,
  %``FeynRules - Feynman rules made easy,''
  arXiv:0806.4194.
  %%CITATION = ARXIV:0806.4194;%%

\bibitem{LopezHonorez:2006gr}
  L.~Lopez Honorez, E.~Nezri, J.~F.~Oliver and M.~H.~G.~Tytgat,
  %``The inert doublet model: An archetype for dark matter,''
  JCAP {\bf 0702}, 028 (2007);
  %%CITATION = JCAPA,0702,028;%%
%\cite{Andreas:2009hj}
%\bibitem{Andreas:2009hj}
 S.~Andreas, M.~H.~G.~Tytgat and Q.~Swillens,
 %``Neutrinos from Inert Doublet Dark Matter,''
 arXiv:0901.1750.
 %%CITATION = ARXIV:0901.1750;%%


%\cite{Bertone:2008xr}
\bibitem{Bertone:2008xr}
  G.~Bertone, M.~Cirelli, A.~Strumia and M.~Taoso,
  %``Gamma-ray and radio tests of the e+e- excess from DM annihilations,''
  arXiv:0811.3744 [astro-ph].
  %%CITATION = ARXIV:0811.3744;%%

%\cite{Cirelli:2008pk}\cite{ArkaniHamed:2008qn}dmdecay
\bibitem{Cirelli:2008pk}
  M.~Cirelli, M.~Kadastik, M.~Raidal and A.~Strumia,
  %``Model-independent implications of the e+, e-, anti-proton cosmic ray
  %spectra on properties of Dark Matter,''
  Nucl.\ Phys.\  B {\bf 813}, 1 (2009)
  [arXiv:0809.2409].
  %%CITATION = NUPHA,B813,1;%%


%
\bibitem{dmdecay}
%\cite{Buchmuller:2007ui}
%\bibitem{Buchmuller:2007ui}
  W.~Buchmuller, L.~Covi, K.~Hamaguchi, A.~Ibarra and T.~Yanagida,
  %``Gravitino dark matter in R-parity breaking vacua,''
  JHEP {\bf 0703}, 037 (2007);
  %%CITATION = JHEPA,0703,037;%%
   C.~R.~Chen, F.~Takahashi and T.~T.~Yanagida,
  %``Gamma rays and positrons from a decaying hidden gauge boson,''
  Phys.\ Lett.\  B {\bf 671}, 71 (2009);
%\cite{Ibarra:2008jk}
%\bibitem{Ibarra:2008jk}
  A.~Ibarra and D.~Tran,
  %``Decaying Dark Matter and the PAMELA Anomaly,''
  arXiv:0811.1555;
  %%CITATION = ARXIV:0811.1555;%%
%\cite{Chen:2008dh}
%\bibitem{Nardi:2008ix}
  E.~Nardi, F.~Sannino and A.~Strumia,
  %``Decaying Dark Matter can explain the electron/positron excesses,''
  JCAP {\bf 0901}, 043 (2009).
    %%CITATION = JCAPA,0901,043;%%




\end{thebibliography}
\end{document}